\documentclass[aps,prb,reprint,superscriptaddress,showpacs,floatfix,longbibliography]{revtex4-2}
\usepackage{amsmath,amssymb,amsfonts,float,graphics,epsfig,epstopdf,color,verbatim,tabularx,bm,multirow,appendix,hyperref}
\pdfoutput=1
\usepackage[normalem]{ulem}
\usepackage{lmodern}
\usepackage{ytableau}
\usepackage{pifont}
\usepackage{color}
\usepackage{bm}
\usepackage{wasysym}
\usepackage{dsfont}
\DeclareMathOperator{\Tr}{Tr}

\def\bk{{\mathbf{k}}}
\def\br{{\mathbf{r}}}
\def\bn{{\mathbf{n}}}

\def\br{{\mathbf{r}}}

\def\bq{{\mathbf{q}}}
\def\bQ{{\mathbf{Q}}}

\def\bS{{\mathbf{S}}}

\begin{document}
	
\title{Quantum Monte Carlo fermion spectroscopy of a non-compact CP$^{1}$ model}

\author{Xu Zhang}
\affiliation{Department of Physics and Astronomy, Ghent University, Krijgslaan 299, 9000 Gent, Belgium}

\author{Nick Bultinck}
\affiliation{Department of Physics and Astronomy, Ghent University, Krijgslaan 299, 9000 Gent, Belgium}

\begin{abstract}
We study a model describing electrons coupled to anti-ferromagnetic spin fluctuations, and consider the situation where hedgehog defects in the order parameter field are suppressed. Without hedgehogs, the bosonic sector of the theory can be taken to realize the physics of the non-compact CP$^1$ theory with a deconfined U$(1)$ gauge field. After strongly coupling the boson to fermion spins, we simulate the single-particle spectral properties of a hedgehog-suppressed electron-boson model defined on a bilayer square lattice with Quantum Monte Carlo, and interpret the results in terms of an effective theory with fractionalized spinon and chargon excitations. As one of our main results we show that the electron gap on top of the half-filled insulator with gapless photon fluctuations closely resembles the mean-field dispersion of an electron in an anti-ferromagnetic spin background, even though the system fully preserves both the translation and spin rotation symmetry. Finally, we discuss potential implications of our results for the high-temperature superconductors.
\end{abstract}
\date{\today}
\maketitle

\section{Introduction}

Metallic phases with non-trivial topological order are of central interest in the study of strongly correlated electron systems. One way to realize such phases is by doping a quantum spin liquid insulator, as first envisioned by Anderson in the context of the resonating valence-bond theory for the cuprate superconductors~\cite{RVB,Baskaran1987,Baskaran1988}. A particularly attractive feature of two-dimensional topological metals is that they can violate the Luttinger constraint on the Fermi surface area \cite{Oshikawa2000,Oshikawa2000_2,Hastings2004,FLstar}, thus allowing for the possibility of small Fermi pockets in the absence of symmetry breaking. This provides a potential mechanism to explain the pseudogap phenomenon in the cuprate phase diagram.

Most metals with topological order evade controlled analytical calculations, as they are described by strongly-coupled lattice gauge theories (one notable exception, however, are models with a $\mathbb{Z}_2$ gauge field which has no quantum fluctuations~\cite{Seifert2018,Coleman2022}). In recent years, a series of impressive numerical works has helped to confirm and further develop the theoretical understanding of topological metals \cite{gazit2017emergent,snir2018confinement,hohenadler2018fractionalized,xu2019monte,wang2019dynamics,janssen2020confinement,Gazit,chen2021fermi,borla2022quantum,raczkowski2022breakdown,brenig2022spinless,borla2024deconfined,pan2025quantum,chen2025emergent}. However, the models studied in these works either (1) use Dirac matter, (2) start from a model with a microscopic gauge field, or (3) do not have full spin rotation symmetry. 

In this work we set out to use Quantum Monte Carlo (QMC) to simulate a metal with an emergent gauge field and SU(2) spin rotation symmetry. Our starting point is a model of electrons with a Fermi surface coupled to an anti-ferromagnetic order parameter field, a model which goes under the name of the spin-fermion model in the high-T$_c$ context~\cite{Monthoux1991,Abanov2003}. To introduce topological order and the associated gauge field into this model we follow Refs.~\cite{lau1989numerical,kamal1993new,motrunich2004emergent}, and sample the O(3) order parameter field over configurations which contain no isolated hedgehog defects. Ref.~\cite{motrunich2004emergent} showed that the hedgehog-free bosonic model defined on a decorated cubic lattice can realize the symmetric phase of the non-compact CP$^1$ model with a deconfined U(1) gauge field and gapped spinon excitations. In this work, we show that the deconfined phase can also be realized on a simple cubic lattice by adding a new term to the lattice Hamiltonian.

By strongly coupling the bosonic field to the spin of fermions hopping on a bilayer square lattice, we obtain an insulating state at half filling in which the fermion spins realize a U(1) spin liquid. Generically, one expects that doping this state will lead to a metal with topological order and small Fermi pockets. However, for our QMC simulations we have to use a bilayer fermion system in order to remove the sign problem, and we find that this setup leads to very strong inter-layer superconductivity upon doping away from half filling -- thus preventing us to directly realize a metallic state. But despite this shortcoming of the model, we are still able to check some predictions made by a \emph{rotating reference frame} theory, which is an effective description in terms of spinon, chargon and U(1) gauge fields~\cite{Shraiman,Schulz,Dupuis,Borejsza,Kaul2007,Kaul2008,Sachdev2009,Chowdhury2015,Chatterjee2017,Sachdev2019,SachdevScammell,Bonetti,Nikolaenko,Vilardi,Forni,Sachdev2025}. In particular, this theory predicts that the dispersion relation for a single hole or electron doped into the U(1) spin liquid insulator is similar to the corresponding mean-field dispersion in an anti-ferromagnet. This means that upon doping one expects small electron or hole pockets at the exact same locations as in a doped anti-ferromagnet, even though there is no symmetry breaking. For the bilayer model used here we find that this theoretical expectation is indeed borne out for the electron dispersion, but not for the hole dispersion (we use a positive next-nearest neighbour hopping to break the particle-hole symmetry).

The main text of this paper is organized as follows. We start in Sec.~\ref{purebos} by introducing the bosonic sector of the theory, and explain how we realize the symmetric phase of the non-compact CP$^1$ model on a simple cubic lattice. In Sec.~\ref{eff} we introduce the complete electron-boson model, and review the rotating frame theory~\cite{Shraiman,Schulz,Dupuis,Borejsza,Kaul2007,Kaul2008,Sachdev2009,Chowdhury2015,Chatterjee2017,Sachdev2019,SachdevScammell,Bonetti,Nikolaenko,Vilardi,Forni,Sachdev2025}. We also provide a theoretical explanation of the strong inter-layer pairing tendency of the model. In Sec.~\ref{counter} we briefly explain how the feedback effect of the fermions on the bosons leads to magnetic order, and how evade this effect in our simulations. Our main numerical results are then presented in Sec.~\ref{elbos}, where we first consider the half-filled case, and then study the situation with small doping. Finally, we summarize our findings in the discussion section \ref{disc}, and comment on potential implications of our results for the cuprate superconductors.

\section{Purely bosonic model}\label{purebos}
We first introduce the bosonic sector, whose low-energy physics should be described by the non-compact CP$^{1}$ model. We start from the well-known (see e.g. \cite{Polyakov}) equivalence between the Non-Linear Sigma Model (NL$\sigma$M) with Euclidean action
\begin{equation}
S = \frac{1}{4g}\int\mathrm{d}\tau \mathrm{d}^2\br\, (\partial_\mu \bn)^2 \label{e1}
\end{equation}
and the CP$^1$ model with action
\begin{equation}
S = \frac{1}{g}\int\mathrm{d}\tau\mathrm{d}^2 \br\,|(\partial_\mu - \mathrm{i} a_\mu)z|^2 \label{e2}.
\end{equation}
Here we have introduced a three-component unit vector field $\bn$, and a corresponding spinon field $z=(z_{\uparrow},z_{\downarrow})^T$ which satisfies $|z|^2\equiv|z_{\uparrow}|^2+|z_{\downarrow}|^2=1$. The connection between $z$ and $\bn$ is $\bn = z^\dagger \boldsymbol{\sigma}z$, where $\boldsymbol{\sigma}$ are the three Pauli matrices. The CP$^1$ theory has a U$(1)$ gauge symmetry and corresponding gauge field $a_\mu$. At the saddle-point level we can identify
\begin{equation}
a_\mu = -\mathrm{i}z^\dagger\partial_\mu z. \label{a}
\end{equation}
With this formula, the Pontryagin density of the NL$\sigma$M maps to the flux density of the CP$^1$ model:
\begin{equation}
\frac{1}{4\pi}\bn\cdot(\partial_\mu \bn \times \partial_\nu \bn) = \frac{1}{2\pi}(\partial_\mu a_\nu - \partial_\nu a_\mu)\,,\label{PF}
\end{equation}
which implies that hedgehog defects of $\bn$ correspond to monopoles of the U$(1)$ gauge field. Our main goal in this section is to realize the physics of the non-compact, i.e. monopole-free, CP$^1$ theory in terms of a three-component unit vector field $\bS$ on a 2+1D Euclidean space-time lattice. In particular, we want to realize a symmetric phase (where spinful excitations are gapped), whose low-energy excitations correspond to the photon of a deconfined U$(1)$ gauge field.

The lattice version of the NL$\sigma$M is a Heisenberg model with partition function
\begin{equation}
	Z_{H} = \int[D \bS]\, \prod_k\delta(\bS_k^2-1)\, e^{\Delta\tau J \sum_{\langle i,j \rangle} \mathbf{S}_i \cdot \mathbf{S}_j},
\end{equation}
where we fix the imaginary time-step to be $\Delta \tau = 0.1$, and always scale the imaginary time direction together with spatial direction $L_\tau=L$. To realize the physics of the non-compact CP$^1$ (NCCP$^1$) model we have to suppress hedgehog defects. To this end we follow Refs.~\cite{berg1981definition,lau1989numerical,kamal1993new,motrunich2004emergent} and introduce a lattice gauge field $a_{ij}$ with orientation $i\rightarrow j$ on the edges of the cubic lattice, which is defined via the parallel transport of a tangent vector on $S^2$ starting from an arbitrarily fixed reference spin $\bS_0$, i.e. $\bS_0\rightarrow\bS_i\rightarrow\bS_j\rightarrow\bS_0$. Denoting the corresponding rotation angle as $\Omega_{0ij}$, we define $a_{ij} = \Omega_{0ij}/2$ mod $2\pi$. Note that $|\Omega_{0ij}|$ is the spherical triangle area spanned by the three spins $\bS_0,\bS_i,\bS_j$. The gauge field can be calculated from the formula
\begin{equation}
	e^{\mathrm{i} a_{ij}}  =\frac{1+\mathbf{S}_0\cdot\mathbf{S}_i+\mathbf{S}_0\cdot\mathbf{S}_j+\mathbf{S}_i\cdot\mathbf{S}_j+\mathrm{i} \mathbf{S}_0\cdot\mathbf{S}_i\times\mathbf{S}_j}{\sqrt{2(1+\mathbf{S}_0\cdot\mathbf{S}_i)(1+\mathbf{S}_0\cdot\mathbf{S}_j)(1+\mathbf{S}_i\cdot\mathbf{S}_j)}}\,,
\end{equation}
Rotating the reference spin from $\bS_0$ to $\bS'_0$, induces a gauge transformation $a_{ij} \rightarrow a_{ij} - \varphi_i + \varphi_j$ with $e^{\mathrm{i}\varphi_i} = e^{\frac{\mathrm{i}}{2}\Omega_{0'0i}}$. 

To a plaquette with sites $(1,2,3,4)$ we can associate a gauge invariant flux $f_{\Box}$
\begin{equation}
	e^{\mathrm{i}f_{\Box}}=e^{\mathrm{i}(a_{12}+a_{23}+a_{34}+a_{41})}\, ,
\end{equation}
where the circulation is defined counter clockwise with respect to the normal vector of the plaquette, which we take to point in the direction of increasing $\tau$, $x$ and $y$. To eliminate the mod $2\pi$ uncertainty, from now on we define $f_\Box \in (-\pi,\pi]$ as a first step towards realizing a non-compact theory. From Eq.~\eqref{PF}, we see that our definition of $a_{ij}$ indeed produces a lattice version of the flux in the CP$^1$ model.

The hedgehog number $Q$ in each elementary cube is given by
\begin{equation}
2\pi Q = \sum_{\Box,+} f_\Box - \sum_{\Box,-} f_\Box\,,
\end{equation}
where the first sum is over the three outward oriented boundary plaquettes, and the second sum is over the three inward oriented boundary plaquettes. The hedgehog number is an integer because it counts how many times the sphere is covered by adding up the solid angles $\pm2f_\Box$. With periodic boundary conditions in all three directions, the total hedgehog number is zero: $\sum_{\text{cubes}}Q=0$. 

To realize the NCCP$^1$ model, isolated hedgehogs should be fully suppressed. We do this in practice by using the same phase space constraint as in Ref.~\cite{motrunich2004emergent}: if $Q\neq 0$ for a cube, there is one and only one nonzero $Q'=-Q$ in the six nearest neighbor cubes. One important difference to Ref.~\cite{motrunich2004emergent}, however, is that we work on a simple cubic lattice, with spins only on the sites of the lattice. At $J=0$, we find that the model is magnetically ordered. To enter the disordered phase we introduce a new term in the Hamiltonian that is designed to frustrate the magnetic order, without changing the general form of the fractionalized description in terms of the CP$^1$ model. Specifically, the partition function we use is
\begin{equation}
	Z_S = \int_{\text{C}}[D \bS]\,\prod_k\delta(\bS_k^2-1)\,e^{-\Delta\tau K\sum_{\Box\perp \tau}f_{\Box}^2}, \label{Z}
\end{equation}
where the integral is over the constrained phase space, and we have put $J=0$ as this is the value that we will use throughout this work. In Eq.~\eqref{Z} the sum is over all spatial plaquettes, i.e. the plaquettes orthogonal to the $\tau$ direction, and $K>0$ corresponds to a lattice version of $(\partial_\mu a_\nu - \partial_\nu a_\mu)^2$ in Maxwell’s Hamiltonian where $\mu,\nu$ take spatial values. As $|f_\Box|$ is half the solid angle spanned by the four spins around a plaquette, positive $K$ should strengthen the ferromagnetic order, while negative $K$ should weaken it. We thus proceed with $K<0$ in our simulations.

\begin{figure}
	\centering
	\includegraphics[width=1\columnwidth]{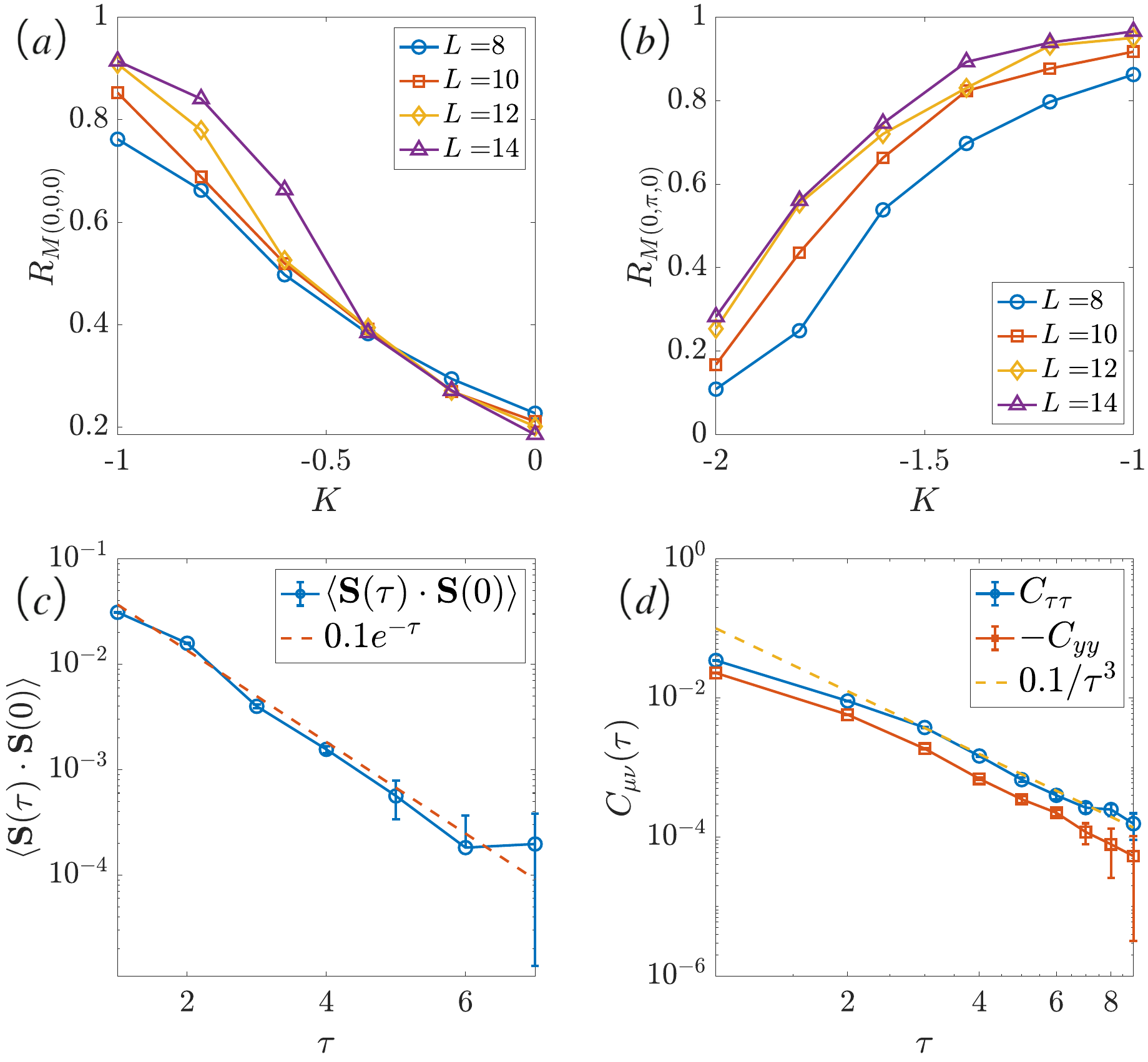}
	\caption{Phase diagram of the purely boson model in Eq.~\eqref{Z} with hedgehog suppression. The ferromagnetic order present around $K=0$ can be seen in (a) and the absence of $\mathbf{q}=(0,\pi,0)$ magnetic order in (b) from the respective correlation ratios $R_{M(0,0,0)}$ and $R_{M(0,\pi,0)}$. These figures also indicate a paramagnetic region around $K=-1$. We fix $K=-1$ for identifying this phase in (c),(d). The spin-gap is confirmed by the exponential decay of the spin correlation function $\langle \mathbf{S}(\tau)\cdot \mathbf{S}(0) \rangle$ along the imaginary time direction as shown in (c). The flux correlations $\langle f^\mu(\tau) f^\mu(0)\rangle$ in (d) with a $1/\tau^3$ decay confirm the Coulomb phase with a gapless photon excitation.}
	\label{fig:Boson_Rmall}
\end{figure}

In Fig.~\ref{fig:Boson_Rmall} we show our numerical results, obtained via local-update Monte Carlo simulations, with additional phase space constraint check for each configuration. To study the magnetic order we have measured the spin structure factor
\begin{equation}
    C_{M}(\mathbf{q})=\frac{1}{L_{\tau}L^2}\sum_{i,j} e^{\mathrm{i}\mathbf{q}\cdot(\mathbf{r}_i-\mathbf{r}_j)}\langle \mathbf{S}_i\cdot \mathbf{S}_j\rangle \,,
\end{equation}
with $L_\tau$ the size in the imaginary-time direction, and $L$ the size of the system in the spatial directions. Using coordinate labels $(\tau,x,y)$, we obtain the magnetic correlation ratios from the structure factor as
\begin{eqnarray}
R_{M(0,0,0)} & = &\frac{C_{M}(\mathbf{q}=(0,0,\Delta q))}{C_{M}(\mathbf{q}=(0,0,0))} \\
R_{M(0,\pi,0)} & = &\frac{C_{M}(\mathbf{q}=(0,\pi,\Delta q))}{C_{M}(\mathbf{q}=(0,\pi,0))}\,,
\end{eqnarray}
where $\Delta q = 2\pi/L$. In a ferromagnetic phase, $R_{M(0,0,0)}$ decreases with increasing $L=L_\tau$, whereas in a paramagnetic phase it increases with system size. At a continuous phase transition between a ferromagnetic and paramagnetic phase $R_{M(0,0,0)}$ is to leading order independent of the system size. In Fig.~\ref{fig:Boson_Rmall} (a) we show the ferromagnetic correlation ratio as a function of $-1< K<0$ for different values of $L=L_\tau$. We see that the ferromagnetic order indeed disappears at a value of $ K\approx -0.4$.

A negative $K$ prefers a solid angle of $2\pi$ between the four spins on every spatial plaquette, and hence could potentially give rise to a coplanar but non-collinear spin order, corresponding to a vortex-anti-vortex checkerboard pattern. However, this pattern cannot be really solidify due to the phase space constraint. To verify this we have calculated the magnetic correlation ratio $R_{M(0,\pi,0)}$ at one of the wavevectors of the checkerboard order, and from Fig.~\ref{fig:Boson_Rmall} (b) we indeed observe no spin order at $\bq = (0,\pi,0)$ even for moderately large $|K|$.

In Fig.~\ref{fig:Boson_Rmall} (c) we show the spin-spin correlation function along the imaginary-time direction obtained with $K=-1$. We see a clear exponential decay, consistent with a paramagnetic phase. To confirm that this paramagnetic phase is described by the NCCP$^1$ theory we have measured following correlation function along the imaginary-time direction
\begin{equation}
C_{\mu\nu}(\tau)=\langle \sin f^\mu(0,0,0) \sin f^\nu(\tau,0,0)\rangle\,, \label{Cmu}
\end{equation}
with $f^\mu(\br_i)$ the flux through the plaquette associated with site $i$ and orthogonal to $\mu$:
\begin{equation}
e^{\mathrm{i}f^\mu(\br_i)} = e^{\mathrm{i} \epsilon^{\mu\nu\sigma}(a_{i,i+\nu} - a_{i+\sigma,i+\sigma+\nu})}
\end{equation}
Identifying this lattice flux as the electric and magnetic fields of a deconfined U$(1)$ gauge theory implies that the correlation function in Eq.~\eqref{Cmu} should decay as
\begin{equation}
C_{\mu\nu}(\br) \sim \frac{3r_\mu r_\nu - \delta_{\mu\nu}r^2 }{r^5}\,,
\end{equation}
and in particular $C_{\tau\tau}(\tau) \sim 2/\tau^3$ and $C_{yy}(\tau) \sim -1/\tau^3$. The numerical results in Fig.~\ref{fig:Boson_Rmall} (d) agree with these expressions, both in terms of the cubic decay and the sign structure. This provides strong evidence that our model in Eq.~\eqref{Z} on the simple cubic lattice indeed realizes the physics of the NCCP$^1$ theory. 

\section{Double layer construction and effective field theory analysis}\label{eff}

In order to obtain a model that is amenable to sign problem-free QMC simulations we introduce fermions on a bilayer square lattice geometry. We write the creation operator for a fermion with spin $\sigma\in\{\uparrow,\downarrow\}$ on site $i$ in layer $l\in\{t,b\}$ as $c^\dagger_{i,l,\sigma}$, and consider the fermion Hamiltonian
\begin{equation}
H_f[\bS] = \sum_{i,j} c^\dagger_i t_{ij}c_j + g\sum_i (-1)^{x_i + y_i} \bS_i \cdot c^\dagger_i l^z \boldsymbol{\sigma} c_i.
\end{equation}
Here $x_i$ and $y_i$ are the integer lattice coordinates of site $i$. The nearest and next-nearest neighbor hopping terms are diagonal in the layer and spin indices, and the coupling to the bosonic field is chosen such that 
\begin{equation}
c_i \rightarrow \mathrm{i} l^x \sigma^y c_i\,\, ,\;\; \mathrm{i}\rightarrow -\mathrm{i} \label{kramers}
\end{equation}
is an anti-unitary Kramers symmetry of $H_f[\bS]$ for every configuration of the $\bS$ field. It is this symmetry which ensures that the system is free of the sign problem in QMC simulations. The staggered nature of the coupling ensures that ferromagnetic order of $\bS$ induces anti-ferromagnetic order for the fermions.

\subsection{Rotating frame chargon theory}
In this section, we discuss an effective field theory which is expected to capture the low-energy physics. We first define the matrix field
\begin{equation}
R_i = \left(\begin{matrix}z_{i,\uparrow} & z^*_{i,\downarrow} \\ z_{i,\downarrow} & -z^*_{i,\uparrow} \end{matrix}\right)\, ,
\end{equation}
which satisfies $R_iR_i^\dagger=R_i^\dagger R_i=\mathds{1}$, and has the property
\begin{equation}
R_i \kappa^z R^\dagger_i = \bS_i\cdot  \boldsymbol{\sigma},\label{Rprop}
\end{equation}
where $\boldsymbol{\kappa}$ are Pauli matrices acting in the column space of $R_i$. Note that spin rotations act with left action
\begin{equation}
R_i \rightarrow U R_i\,,
\end{equation}
whereas U$(1)$ gauge transformations $z_i \rightarrow e^{\mathrm{i}\theta_i}z_i$ are implemented with a right action
\begin{equation}
R_i \rightarrow R_i e^{\mathrm{i}\theta_i \kappa^z}.
\end{equation}
Using Eq.~\eqref{Rprop} the fermionic part of the action is
\begin{eqnarray}
S_f & = & \int\mathrm{d}\tau \bigg[ \sum_{i,j} \bar{c}_i(\delta_{ij} \partial_\tau  + t_{ij} ) c_j  \\
 & & \hspace{1 cm} + g\sum_i (-1)^{x_i+y_i} \bar{c}_i R_i l^z \kappa^z R^\dagger_i c_i \bigg] \nonumber
\end{eqnarray}
We can now introduce new gauge-dependent fermion fields \cite{Shraiman,Schulz,Dupuis,Borejsza,Kaul2007,Kaul2008,Sachdev2009,Chowdhury2015,Chatterjee2017,Sachdev2019,SachdevScammell,Bonetti,Nikolaenko,Vilardi,Forni,Sachdev2025}
\begin{eqnarray}
\bar{\psi}_i & = & \bar{c}_iR_i \\
\psi_i & = & R_i^\dagger c_i\, ,
\end{eqnarray}
whose components $\bar{\psi}_{i,+} = \bar{c}_i z_i$ and $\bar{\psi}_{i,-} = \bar{c}_i \mathrm{i}\sigma^y z^*_i$ are spin-singlet and have U$(1)$ gauge charge $\pm 1$. In terms of these new chargon fields we can rewrite the action as 
\begin{eqnarray}
S_f & = & \int\mathrm{d}\tau \bigg[ \sum_{i,j} \bar{\psi}_i(\delta_{ij} \partial_\tau  + t_{ij} R^\dagger_i R_j) \psi_j  \\
&& \hspace{0.4 cm} + \sum_i \bar{\psi}_i(R_i^\dagger\partial_\tau R_i)\psi_i 
 + g\sum_i (-1)^{x_i+y_i} \bar{\psi}_i l^z \sigma^z \psi_i \bigg].\nonumber
\end{eqnarray}
The physical interpretation of this action can be made more transparent by noting that
\begin{eqnarray}
R_i^\dagger \partial_\tau R_i & = &  \kappa^z z^\dagger_i\partial_\tau z_i + \left(\begin{matrix}  0 & z^\dagger_i\mathrm{i}\sigma^y\partial_\tau z^*_i\\ \partial_\tau z^T_i\mathrm{i}\sigma^y z_i & 0 \end{matrix} \right) \label{Rtau}\\
R_i^\dagger R_j & = & |z^\dagger_i z_j| e^{\mathrm{i} a_{ij}\kappa^z} + \left(\begin{matrix}  0 & z^\dagger_i\mathrm{i}\sigma^y z^*_j\\ -z^T_i\mathrm{i}\sigma^y z_j & 0 \end{matrix} \right)  \label{Rij}
\end{eqnarray}
In Eq.~\eqref{Rtau} we recognize the temporal component of the gauge field $z_i^\dagger\partial_\tau z_i = \mathrm{i}a_0$ according to Eq.~\eqref{a}, and in Eq.~\eqref{Rij} we have introduced the notation arg$(z_i^\dagger z_j) = a_{ij}$ as suggested by the gradient expansion $z_i^\dagger z_{i+\mu} \approx  1 + z_i^\dagger\partial_\mu z_i = 1 + \mathrm{i} a_\mu \approx e^{\mathrm{i}a_\mu}$.
We can thus write the fermionic part of the action as
\begin{eqnarray}
S_f & = & \int\mathrm{d}\tau \bigg[ \sum_{i,j} \bar{\psi}_i(\delta_{ij} \left[\partial_\tau +  \mathrm{i}a_{0} \kappa^z\right] + \tilde{t}_{ij} e^{\mathrm{i}a_{ij}\kappa^z}) \psi_j \nonumber \\
&& \hspace{1 cm} 
 + g\sum_i (-1)^{x_i+y_i} \bar{\psi}_i l^z \kappa^z \psi_i \bigg] + \dots \, ,
\end{eqnarray}
where $\tilde{t}_{ij}$ are effective renormalized hopping amplitudes, and the dots represent additional terms coming from the off-diagonal elements in Eqs. \eqref{Rtau} and \eqref{Rij}. These correspond to spinon-chargon coupling terms which for our high-level analysis here are not important. 

We see that the $\psi$ fermions, which live in a rotating spin reference frame that locally follows the fluctuations of the $\bS$ field \cite{Shraiman,Schulz,Dupuis,Borejsza,Kaul2007,Kaul2008,Sachdev2009,Chowdhury2015,Chatterjee2017,Sachdev2019,SachdevScammell,Bonetti,Nikolaenko,Vilardi,Forni,Sachdev2025}, experience a static staggered potential of strength $g$. As a result, the free chargon energies are
\begin{equation}
E_{\mathbf{k}}^{1,2} = \frac{1}{2}(\varepsilon_{\bk} + \varepsilon_{\bk+\bQ}) \pm \frac{1}{2}\sqrt{(\varepsilon_{\bk} - \varepsilon_{\bk+\bQ})^2 + 4g^2}\, ,
\end{equation}
with $\varepsilon_{\bk}$ the dispersion generated by $\tilde{t}_{ij}$ and $\bQ=(\pi,\pi)$. The free-chargon energies are the same for the chargons with positive and negative gauge charge. Importantly, at half filling the free-chargon spectrum is gapped. In the symmetric phase of the NCCP$^1$ model the spinons are also gapped, and hence at half filling we expect the low-energy fluctuations correspond to the photon of the deconfined U$(1)$ gauge field. We emphasize that this insulating state preserves both the spin rotation symmetry and the single-site translation symmetry. This is because translation acts on the chargons as $\bar{\psi}_i \rightarrow \kappa^y \bar{\psi}_{i+x/y}$, and this action leaves the staggered chargon potential invariant. Nevertheless, the dispersion of chargons doped into the insulating state is identical to the mean-field dispersion of an electron or hole doped into an anti-ferromagnetic insulator on the square lattice. Here we have of course assumed that the gauge interaction does not significantly modify the chargon dispersion. This assumption cannot be justified, however, as the gauge field is strongly coupled to the chargons and spinons. 

\subsection{Inter-layer pairing}

Doping a single electron into the insulating state introduces both a spinon and a chargon. Again ignoring the gauge interaction, the electron gap is thus $|g|+m$, where $m$ is the spinon mass. On the other hand, the energy of an electron pair 
\begin{equation}
c^\dagger_{i}l^x\sigma^y c^\dagger_{i} = -\psi^\dagger_{i}l^x\kappa^y \psi^\dagger_{i} \label{cp}
\end{equation}
on top of the half filling state is only $\sim 2|g|$, as in this case no spinons are created. The binding energy of the inter-layer spin singlet pair is thus $\sim 2m$. The energy of a spinon-chargon pair can potentially be lowered by forming a bound state due to the gauge interaction, but when the spinon mass is sufficiently large the energy of a spin-singlet pair should be lower than the energy of two isolated spinon-chargon pairs. As a result, upon doping the symmetric insulator we expect a finite density of tightly-bound inter-layer Cooper pairs, which condense and give rise to superconductivity. For the sign problem-free model studied in this work one can even use the Kramers symmetry in Eq.~\eqref{kramers} to prove that superconductivity is one of the leading ordering instabilities, and that it must occur in the inter-layer spin-singlet channel of Eq.~\eqref{cp} \cite{zhang2025constraints}.

The local pairing mechanism does not depend on the long-wavelength photon fluctuations. In fact, it can be seen to survive deep in the confined phase when we add back the monopoles. In terms of the vector field $\bS$ the confined phase corresponds to the standard disordered phase. In the extreme case where the vector field has a zero space-time correlation length, we can trivially integrate it out and find
\begin{align}
&\frac{1}{4\pi} \int \mathrm{d} \bS_i(\tau)\,e^{\pm \Delta \tau g\, \bS_i(\tau)\cdot c^\dagger_i(\tau) l^z \boldsymbol{\sigma}c_i(\tau)} \\
&= 1 + \frac{1}{4\pi}\int \mathrm{d} \bS_i(\tau)\, \frac{(\Delta \tau g)^2}{2 }(\bS_i(\tau)\cdot c^\dagger_i(\tau) l^z \boldsymbol{\sigma}c_i(\tau))^2 + \dots \nonumber \\
& = 1 +  \frac{1}{3}\frac{(\Delta \tau g)^2}{2 }( c^\dagger_i(\tau) l^z \boldsymbol{\sigma}c_i(\tau))^2 + \dots \nonumber \\
& \approx e^{\frac{(\Delta \tau g)^2}{6} ( c^\dagger_i(\tau) l^z \boldsymbol{\sigma}c_i(\tau))^2}
\end{align}
The boson fluctuations thus induce an attractive fermion interaction of the form
\begin{align}
& -\frac{\Delta \tau g^2}{6} ( c^\dagger_i l^z \boldsymbol{\sigma}c_i)^2 \\
& \hspace{0.5 cm} = \frac{\Delta \tau g^2}{3}\left[ 4 \boldsymbol{s}_{i,t}\cdot \boldsymbol{s}_{i,b}  +\frac{3}{2}\sum_l n_{i,l}(n_{i,l}-2)\right] \, , \nonumber
\end{align}
where $n_{i,l}, \boldsymbol{s}_{i,l}=\frac{1}{2}c^{\dagger}_{i,l}\boldsymbol{\sigma}c_{i,l}$ are electron density and spin operators on site $i$ in layer $l$. The induced anti-ferromagnetic interaction favors the formation of inter-layer spin singlets of the form in Eq.~\eqref{cp}. At half filling, these singlet pairs form a Bose insulator if the hopping is sufficiently small. Upon doping the preformed Cooper pairs become mobile and superconductivity results. In the above analysis we have assumed a finite imaginary time discretization, which is the situation that applies to our QMC simulations presented below.

The fact that monopole proliferation can lead to a featureless Bose insulator of singlet pairs is possible because the monopole Berry phases cancel between the top and bottom fermion layers. This property is intimately related to the sign problem-free nature of the model -- both can be traced back to the $l^z$ matrix in the fermion-boson coupling, which causes the fermion spins to be oriented oppositely in both layers. 

\section{Boson counterterm}\label{counter}

In the previous section we have assumed that the boson field $\bS$ realizes the symmetric phase of the NCCP$^1$ model, even after coupling to the fermions. This is not guaranteed to be true, because of the feedback of the fermions on the bosonic degrees of freedom. Concretely, the effective action for the boson field is $S_b[\bS] - \ln Z_f[\bS]$, with $S_b$ the bare boson action and $Z_f[\bS]$ the free fermion partition function which depends on the space-time configuration of $\bS$. 

To understand the effect of the free fermion partition function, let us examine the half-filling case with $t_{ij}=0$, such that the fermions are localized and behave as spin-$1/2$ degrees of freedom in a time-dependent magnetic field $\pm \bS$ (depending on the layer). Further details are provided in the appendix, but let us note here that on a discretized imaginary-time grid every spin is described by the well-known action (see e.g. \cite{FradkinBook})
\begin{align}
& -   \sum_\tau  \frac{1}{2} \ln\left(\frac{1+\bn(\tau+\Delta \tau)\cdot\bn(\tau)}{2} \right) \label{spinact}\\
& \hspace{1 cm}\pm \sum_\tau (\Delta \tau g) \bS(\tau) \cdot \bn(\tau) - \mathrm{i}S_B[\bn]  \, , \nonumber 
\end{align}
where $\bn$ describes the direction of the fermion spin-$1/2$ in the spin coherent state basis, $\mathrm{i}S_B[\bn]$ is the Berry phase term, and the $\pm$-sign depends on the sublattice and layer. At strong coupling, the direction of the fermion spin-$1/2$ will be aligned (or anti-aligned depending on the layer and sublattice) with the boson field $\bS$. This in turn generates a temporal stiffness term for the boson field $\bS$, which is inherited from the logarithmic term in Eq.~\eqref{spinact}. We thus find that even in the absence of hopping terms the fermions have a non-trivial feedback effect on the boson field which enhances magnetic order. In our numerical simulations we indeed find that coupling the $\bS$ degrees of freedom to dispersionless fermions at half filling destroys the symmetric phase of the NCCP$^1$ model, which gives way to ferromagnetic order (and hence anti-ferromagnetic order for the fermion spins). 

To remedy this, we add a counterterm to $S_b[\bS]$ which is designed to cancel the fermion contribution to the effective boson action. It is important to note that physically allowed counterterms should be local and real. The locality condition is satisfied when the fermions are gapped, which is clearly true for the dispersionless case discussed above. For our model it will also remain true with non-zero hopping and upon doping away from half filling, where the fermion spectrum has a superconducting gap. 

The ability to cancel the fermion contribution to the effective action with a real counterterm is exactly what makes the model sign problem-free. So even though we will sample the boson field $\bS$ according to exactly the same probability distribution as in Sec.~\ref{purebos}, i.e. without the fermion partition function, the fact that the fermion partition function is real is still a crucial property of the theory -- it ensures that we are simulating a physically sensible theory, i.e. a theory for which we can cancel the fermion partition function with a physically allowed bosonic counterterm.

\section{Numerical results for the electron-boson model}\label{elbos}

In this section we present our numerical QMC results for the electron-boson model introduced in the previous section. We first focus on half filling, and then consider the situation with small doping.

\subsection{Half filling and dispersion of single electron or hole}

\begin{figure}
	\centering
	\includegraphics[width=1\columnwidth]{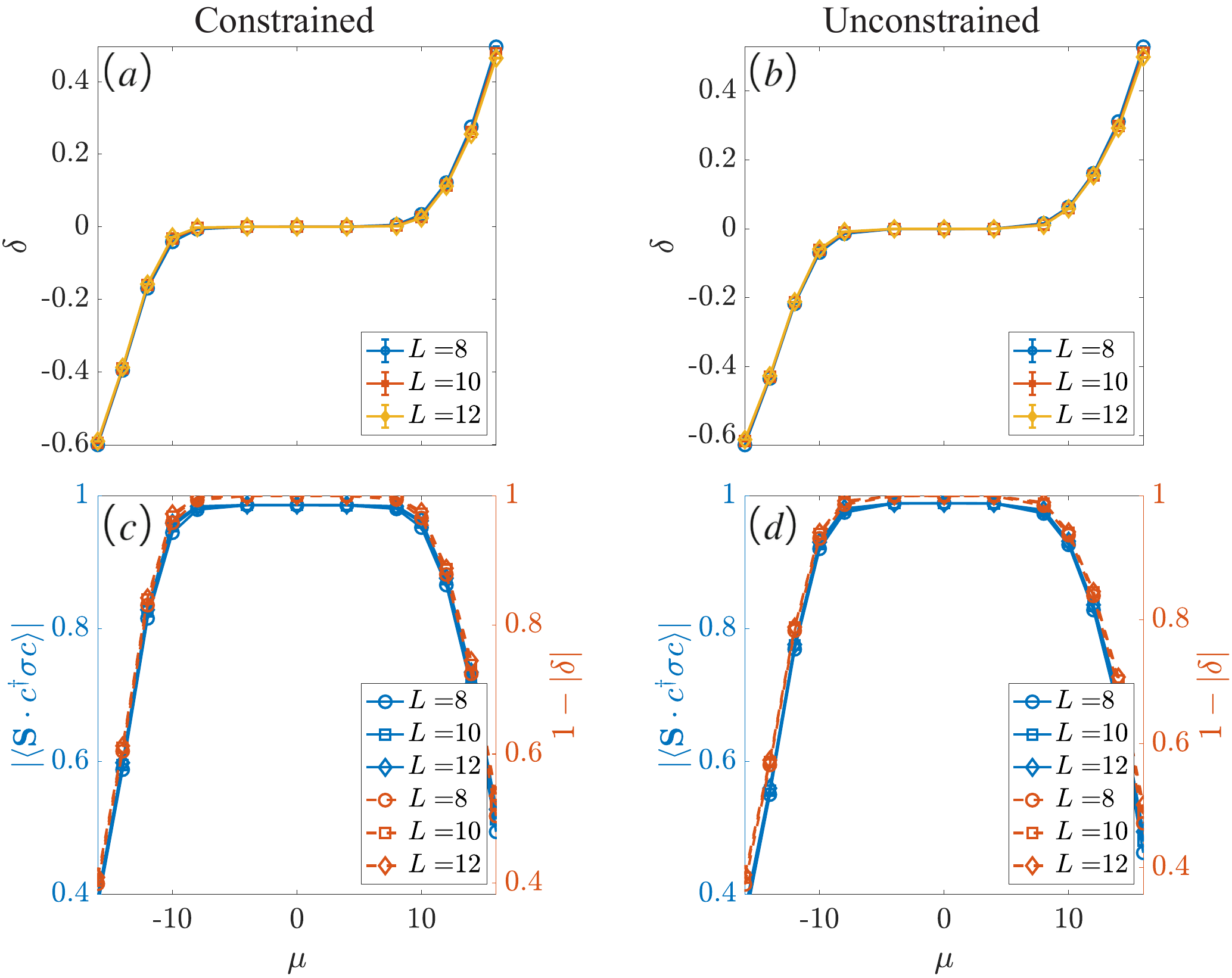}
	\caption{Fermion phases at strong boson-fermion coupling $g=20$ with $K=-1$ for constrained/unconstrained (left/right columns) cases. The charge gap is confirmed from the plateau in the doping density $\delta$ as a function of chemical potential, both for constrained (a) and unconstrained (b) cases. The local alignment between the fermion spin and the boson field is confirmed from $|\langle \mathbf{S}_i \cdot c^{\dagger}_i \mathbf{\sigma}c_i \rangle|\approx 1 - |\delta |$ as shown in (c)(d). }
	\label{fig:Fermion_Rsall}
\end{figure}

\begin{figure*}
	\centering
	\includegraphics[scale=0.4]{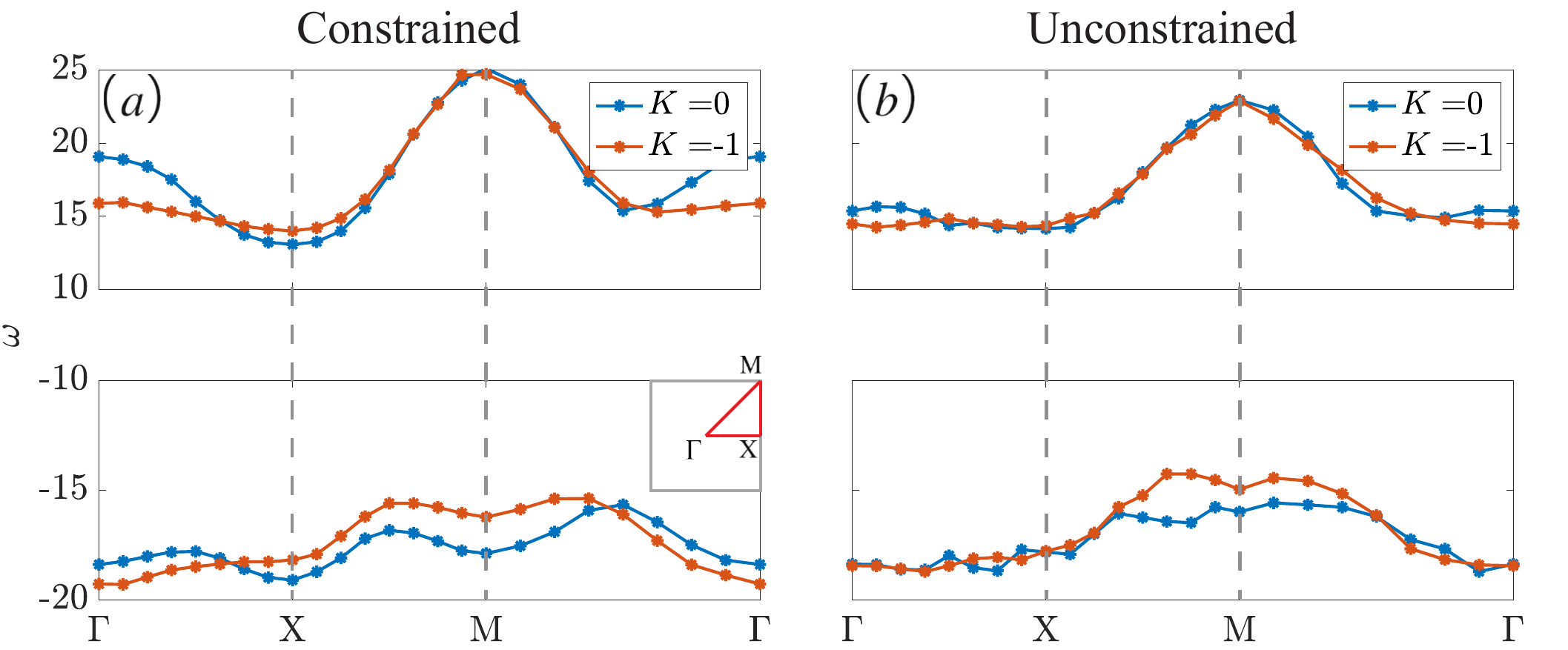}
	\caption{Electron and negative hole gaps obtained from the imaginary time Green's function at strong boson-fermion coupling $g=20$ and hopping parameters $t=-3,t'=-0.35t$ for system size $L=L_\tau = 16$. (a) Results obtained with constrained (i.e. hedgehog-free) sampling. (b) Results obtained with unconstrained sampling. In (a), $K=0$ realizes the AFM phase, while $K=-1$ realizes the spin-gapped topological phase. In (b), both $K=0$ and $K=-1$ correspond to a spin-gapped featureless state.}
	\label{fig:SG_all}
\end{figure*}

\begin{figure}
	\centering
	\includegraphics[width=1\columnwidth]{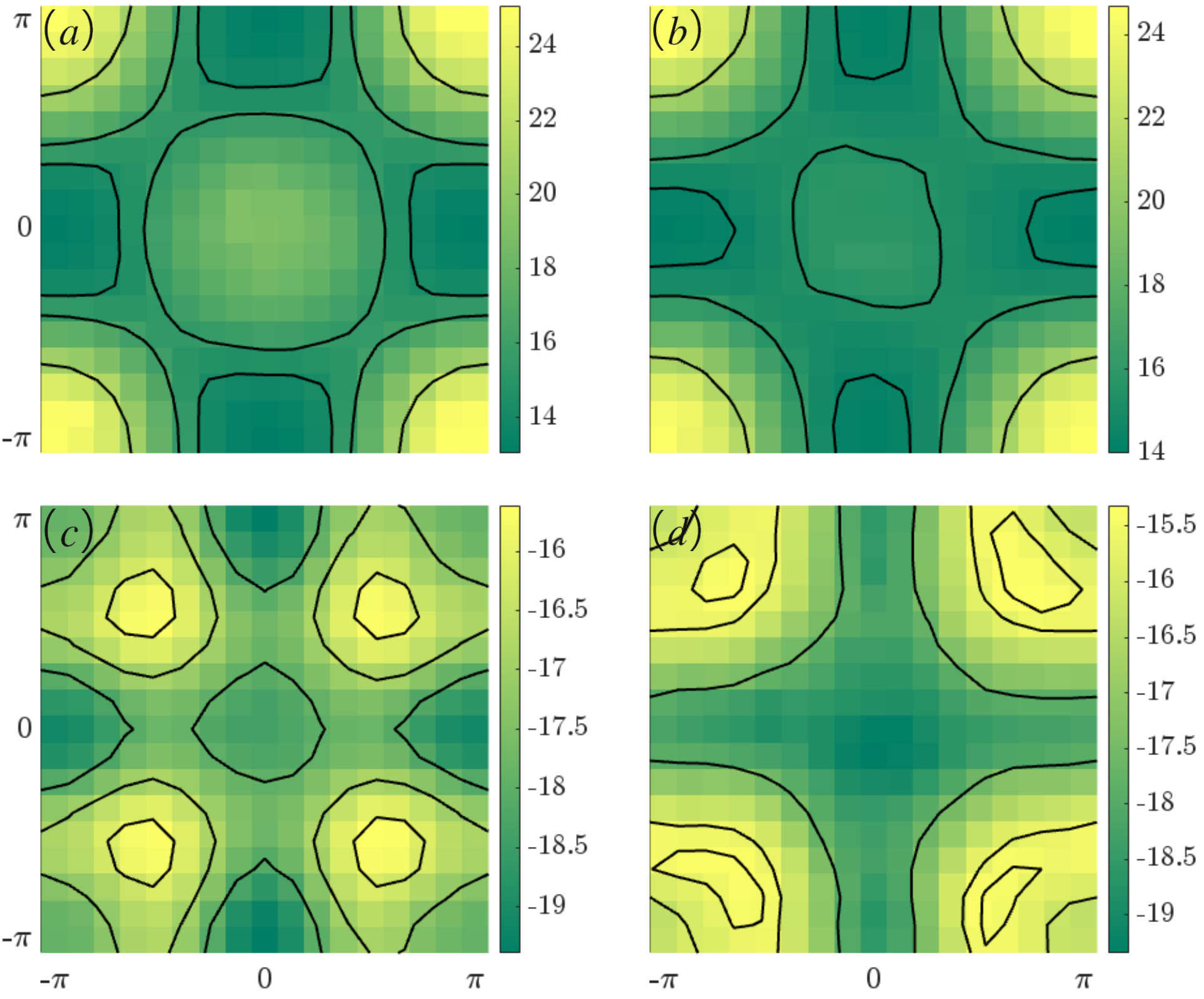}
	\caption{Electron and negative hole gaps obtained from the imaginary time Green's function at strong boson-fermion coupling $g=20$ and hopping parameters $t=-3,t'=-0.35t$. Constrained (i.e. hedgehog-free) sampling is used. (a)(c) show conduction and valence bands (respectively electron and negative hole gaps) at $K=0$; (b)(d) are conduction and valence bands obtained at $K=-1$. The color bar indicates the fitting gap value. System size is $L=L_\tau = 16$.}
	\label{fig:CVBall}
\end{figure}

To obtain an insulator at half filling we work at strong electron-boson coupling $g \gg t$. Specifically, we will use values of $t=-2,-3$ for the nearest-neighbor hopping, fix the next-nearest neighbor hopping as $t' = -0.35 t$, and use coupling constants $g=20$ and $g=10$ (with results for the smaller coupling constant presented in the appendix).

First we confirm that the half-filling state is incompressible by calculating the doping density $\delta = 1 - n$, with $n$ the electron density per site per layer, as a function of the chemical potential $\mu$ (note $\delta$ is defined to be positive for hole doping). In Fig.~\ref{fig:Fermion_Rsall} (a) we see a clear plateau in the $\delta-\mu$ curve located at $\delta = 0$, showing that the compressibility is indeed zero at half filling. The results in Fig.~\ref{fig:Fermion_Rsall} (a) were obtained with the constrained phase space sampling for the boson field $\bS$, i.e. by excluding the hedgehogs in order to realize the physics of the NCCP$^1$ model. We used $K=-1$, which as we confirmed in Sec.~\ref{purebos} realizes the symmetric phase of the NCCP$^1$ model with gapless photon fluctuations. In Fig.~\ref{fig:Fermion_Rsall} (b) we show $\delta$ as a function of $\mu$ for the unconstrained case, i.e. with unrestricted phase space sampling for $\bS$ (again using $K=-1$). Also in the presence of hedgehogs the half-filled state is incompressible. 

The difference between the insulating states obtained without and with hedgehogs should be in their spin properties. In Figs.~\ref{fig:Fermion_Rsall} (c)(d), we plot $|\langle \mathbf{S}_i \cdot c^{\dagger}_{i,t} \mathbf{\sigma}c_{i,t} \rangle |$ to confirm that the fermion spins in the top layer are aligned or anti-aligned with the boson field $\bS$. To very good accuracy we find that $|\langle \mathbf{S}_i \cdot c^{\dagger}_{i,t} \mathbf{\sigma}c_{i,t} \rangle | \approx 1- |\delta |$, showing that there are almost no doubly occupied (empty) sites for hole (electron) doping, and that every occupied fermion site produces a spin-$1/2$ that is almost perfectly aligned or anti-aligned with $\bS$. This implies that in the constrained case, the algebraic flux correlations are imprinted on the fermion spins, and in the constrained case the fermion spins are short-range correlated. So integrating out the constrained boson field produces two perfectly correlated U$(1)$ spin liquids on the fermion layers. Without constraint we obtain a featureless state that is adiabatically connected to a Bose insulator of inter-layer spin-singlet pairs. 

To obtain information on the momentum-dependence of the gap to electron and hole excitations we have measured the imaginary-time Green's function
\begin{equation}
G(\bk,\tau) = \langle c_{\bk,t}(\tau) c^\dagger_{\bk,t}(0) \rangle_{\beta}\, , 
\end{equation}
where $c^\dagger_{\bk,t},c_{\bk,t}$ creates or annihilates an electron with crystal momentum $\bk$ in the top layer, and $\langle\cdot \rangle_\beta$ is the thermal expectation value at inverse temperature $\beta$. By fitting $\tau\gtrsim 0$ part of the Green's function to an exponential decaying function we can obtain an estimate of the electron-excitation gap at different momenta, and by fitting $\tau\lesssim\beta$ part we obtain an estimate for the hole-excitation gap.

\begin{figure}
	\centering
	\includegraphics[width=1\columnwidth]{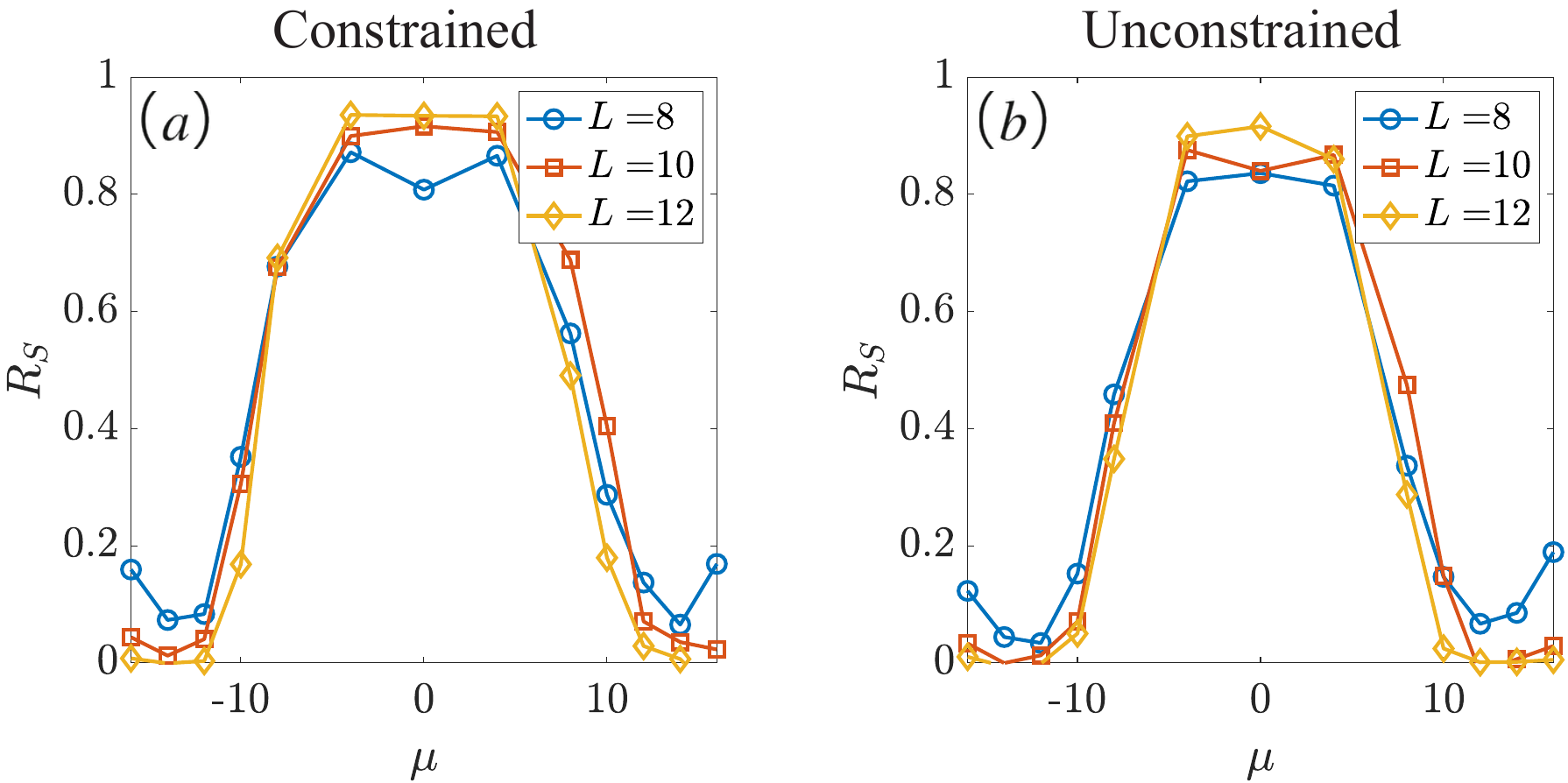}
	\caption{Superconducting correlation ratio $R_S$ as a function of chemical potential $\mu$, obtained with $g=20$, $t = -2$, and $K=-1$. (a) Results with unconstrained sampling. (b) Results with constrained sampling. }
	\label{fig:Fermion_SC}
\end{figure}

\begin{figure*}
	\centering
	\includegraphics[scale=0.3]{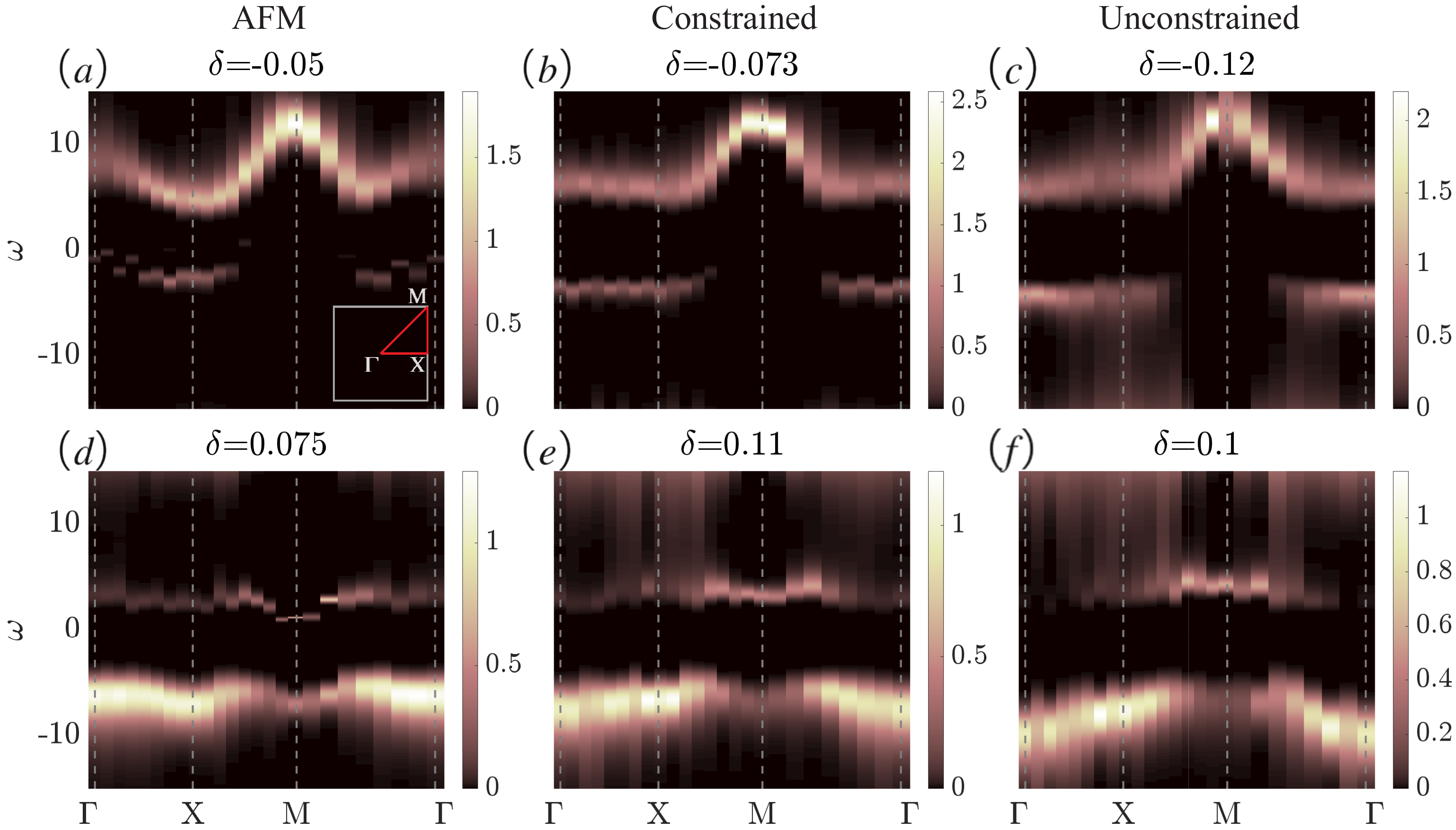}
	\caption{Single-particle spectral function along a high-symmetry line in the Brillouin zone from stochastic analytical continuation (SAC)~\cite{sandvik1998stochastic,beach2004identifying,shao2023progress} of the imaginary-time Green's function obtained for the doped magnetic (a,d), topological (b,e) and featureless (c,f) insulators. Top row shows electron doping $\delta <0$, and bottom row shows hole doping $\delta >0$. The magnetic insulator is obtained with constrained sampling at $K=0$, the topological insulator with constrained sampling at $K=-1$, and the featureless insulator with unconstrained sampling at $K=-1$. The parameters used are $t = -2$ and $g = 20$. System size is $L=L_\tau = 16$.}
	\label{fig:SWall}
\end{figure*} 

The results of this procedure, obtained at $t=-3$ and $g=20$ for a system of size $L=L_\tau = 16$, are shown in Figs.~\ref{fig:SG_all} and \ref{fig:CVBall}. In Fig.~\ref{fig:SG_all} (a) we show the electron gap (top) and negative hole gap (bottom) along a cut through the Brillouin zone connecting high-symmetry points, for the constrained cases both with $K=0$ and $K=-1$. At $K=0$ (blue curves), the system is magnetically ordered, and the momentum-dependent gaps reproduce the general features of the mean-field electron and hole dispersion on top of an anti-ferromagnetic insulator on the square lattice: the electron band has minima at $(\pi,0)$ and $(0,\pi)$ (the X points), and the hole band has maxima at ($\pm \pi/2,\pm \pi/2)$, i.e. right in the middle of the lines connecting $(0,0)$ and $(\pi,\pi)$ (respectively $\Gamma$ and M). With $K=-1$, the system is in the topological phase with algebraic flux correlations. From the simple considerations in Sec.~\ref{eff} we would expect an electron and hole gap following the chargon dispersion, which in turn is expected to resemble the mean-field dispersion in the presence of anti-ferromagnetic order. From the top panel in Fig.~\ref{fig:SG_all} (a) we see that this expectation is indeed borne out for the electron excitations: the electron gap has minima at the $X$ points and (local) maxima at $\Gamma$ and $M$. The most important difference between the electron dispersions on top of the magnetic and topological states is in the height of the local maximum at $\Gamma$: in the topological case the gap at $\Gamma$ is significantly smaller. On the other hand, the hole gap in the topological state deviates significantly from that in the anti-ferromagnetic case. In particular, we observe an additional global band maximum along the X-M line which is not present in the magnetically ordered case.

For reference we also show the electron and (negative) hole gaps on top of the featureless insulator obtained with unconstrained boson sampling in Fig.~\ref{fig:SG_all} (b). Both gaps with $K=0$ and $K=-1$ are very different from anti-ferromagnetic case. The electron band has become much flatter and shows no minima at X, and the hole band has no global maxima on the $\Gamma$-M line. 

In Fig.~\ref{fig:CVBall} we show colorplots of the electron and negative hole gaps in the entire Brillouin zone, obtained with constrained sampling. On these colorplots we added equal-energy contours as a guide to the eye. Figs.~\ref{fig:CVBall} (a,b) show the electron gaps for the anti-ferromagnetic (a) and topological state (b). Again we see good agreement between both in the broad features: (local) maxima at $\Gamma$ and M, and global minima at X. Figs.~\ref{fig:CVBall} (c) and (d) show the negative hole gaps. In Fig.~\ref{fig:CVBall} (c) we recognize the band maxima at $(\pm \pi/2,\pm \pi/2)$ characteristic of the hole dispersion in an anti-ferromagnet. For the negative hole gap on top of the topological state in Fig.~\ref{fig:CVBall} (d) we see that the band maxima have moved away from $(\pm \pi/2,\pm \pi/2)$.

\begin{figure*}
	\centering
	\includegraphics[scale=0.3]{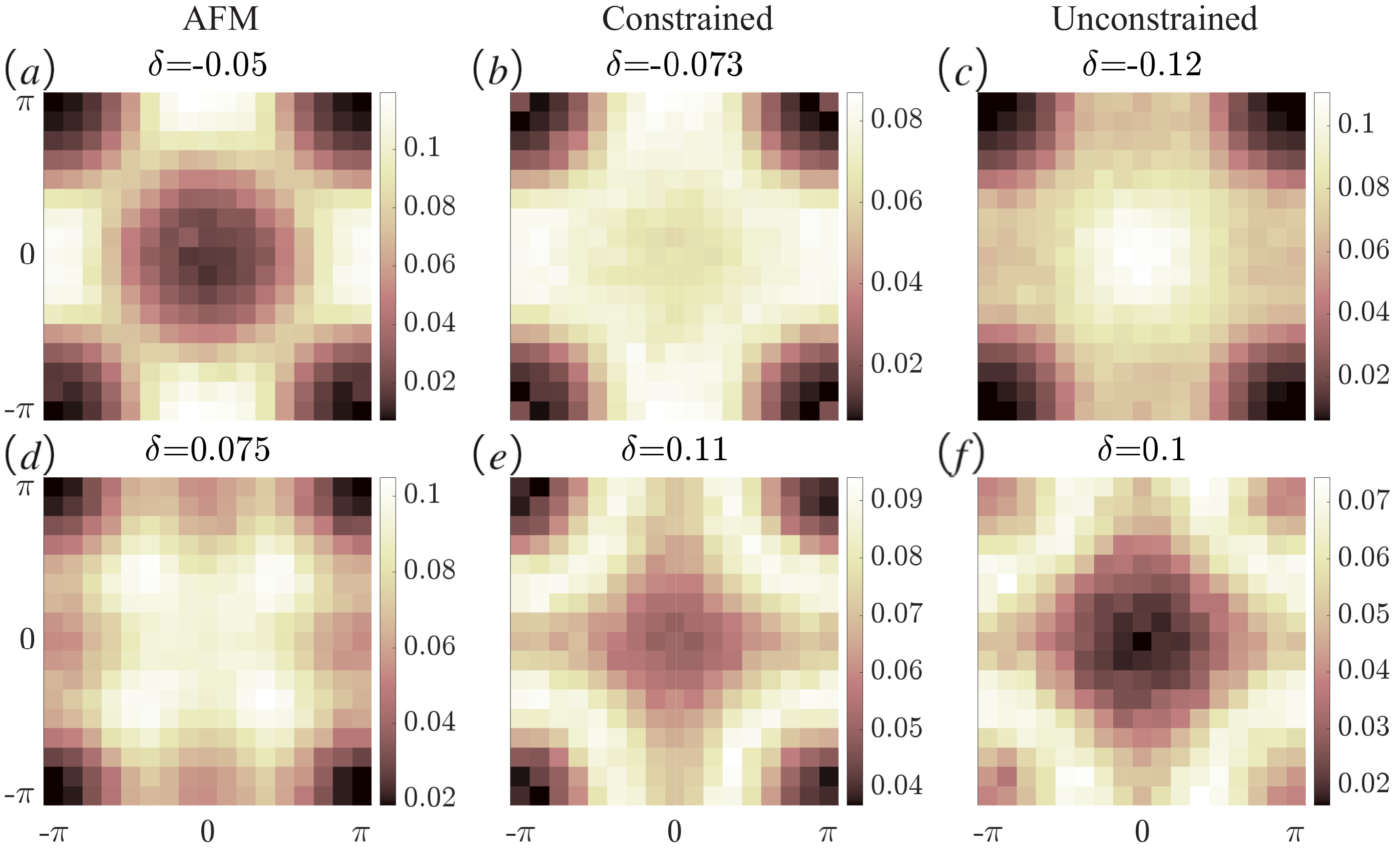}
	\caption{Temperature-broadened zero-frequency spectral weight $W(\bk)$ (as defined in Eq.~\eqref{defW}) over the entire Brillouin zone for the doped magnetic (a,d), topological (b,e) and featureless (c,f) insulators. Top row shows electron doping $\delta <0$, and bottom row shows hole doping $\delta >0$. The magnetic insulator is obtained with constrained sampling at $K=0$, the topological insulator with constrained sampling at $K=-1$, and the featureless insulator with unconstrained sampling at $K=-1$. The parameters used are $t = -2$ and $g = 20$. System size is $L=L_\tau = 16$.}
	\label{fig:FSall}
\end{figure*}

\subsection{Finite doping}

We now turn to the case with small doping density $\delta$. To identify the superconductors which result from doping away from half filling we measured the static superconducting structure factor
\begin{equation}
    C_{S}(\mathbf{q})=\frac{1}{L^2}\sum_{i,j} e^{\mathrm{i}\mathbf{q}\cdot(\mathbf{r}_i-\mathbf{r}_j)}\langle \Delta^\dagger_i\Delta_j\rangle \,,
\end{equation}
where $i,j$ are summed over all sites of the square lattice, $\Delta_i = c_i l^x \sigma^y c_i$, and $\mathbf{r}_i = (x_i,y_i)$ are the spatial coordinates of site $i$. From this structure factor we obtain the superconducting correlation ratio
\begin{equation}
R_S=\frac{C_{S}(\mathbf{q}=(0,\Delta q))}{C_{S}(\mathbf{q}=(0,0))}
\end{equation}
which decreases with system size in a superconducting state, and increases when superconductivity is absent. In Fig.~\ref{fig:Fermion_SC} we show $R_S$ as a function of $\mu$ for different values of $L=L_\tau$, both for the case with constrained and unconstrained boson sampling. We see that for any non-zero doping $\delta$, the system becomes superconducting. This holds for both electron ($\delta < 0$) and hole ($\delta > 0$) doping of the topological as well as the featureless insulator. 

As explained in Sec.~\ref{eff}, we can understand superconductivity in the doped featureless insulator from the liberation of pre-formed inter-layer Cooper pairs. In the topological state, the analysis of Sec.~\ref{eff} suggests an on-site inter-layer pairing with a binding energy determined by the spinon mass. In both cases we thus expect the superconductors to be in the Bose-Einstein-Condensation (BEC) regime. To confirm this we have calculated the single-particle spectral function
\begin{equation}
A(\bk,\omega) = -\frac{1}{\pi}\text{Im}G(\bk,\omega-\mathrm{i}\epsilon)
\end{equation}
via stochastic analytical continuation (SAC)~\cite{sandvik1998stochastic,beach2004identifying,shao2023progress} of the imaginary-time Green's function. In Fig.~\ref{fig:FSall} we show the single-particle spectral weight for different doping values in the range $0.05 \leq |\delta | \leq 0.12$. We see that even at non-zero doping the Fermi energy does not cut through the electron or hole bands, which means that the doped charges have entered as pairs, as expected for a BEC superconductor. This is true not only for doping the topological and featureless insulators, but also for the magnetic insulator.

We further computed the zero-frequency single-particle spectral with finite-temperature broadening:
\begin{equation}
W(\bk):=\frac{\beta}{2}\int \mathrm{d}\omega  \frac{ A(\bk,\omega)}{\cosh(\beta \omega)+1} \label{defW}
\end{equation}
The result using $\beta = 0.5$ is shown in Fig.~\ref{fig:FSall} for the electron and hole doped magnetic, topological and featureless insulators. From these plots we again see the characteristic features of the electron and hole bands on top of the different insulators. In particular, for electron (hole) doping, the broadened zero-frequency spectral weight will be primarily determined by the electron (hole) dispersion relation. The zero-frequency spectral weights confirm that the electron bands on top of the magnetic and topological insulators share important features such as positions of dispersion minima and maxima, whereas the hole band on top of the topological insulator looks more similar to the hole band of the featureless insulator, as it lacks the characteristic maxima at $(\pm \pi/2,\pm \pi/2)$.

\section{Discussion}\label{disc}

In this work we have shown how to realize the symmetric phase of the NCCP$^1$ model, characterized by a deconfined U(1) gauge field, on a simple cubic lattice by sampling over hedgehog-free configurations in an O(3) Heisenberg model. By strongly coupling these Heisenberg spins to a bilayer square lattice of fermions we realized an incompressible state at half-filling where the fermion spins form a U(1) spin liquid. From the decay of the imaginary-time Greens' function, and from the single-particle spectral weight produced by stochastic analytic continuation, we obtained the dispersion relation of a single electron or hole doped in this U(1) spin liquid. The electron dispersion relation is found to be of the same form as the mean-field electron dispersion in the presence of anti-ferromagnetic order, as expected from an effective rotating-frame chargon theory. The hole dispersion in the spin liquid, on the other hand, is found to differ substantially from the mean-field hole dispersion in an anti-ferromagnet. 

We note that doping an electron or hole into the incompressible spin liquid corresponds to simultaneously injecting both a chargon and a spinon, which interact strongly via the emergent U(1) gauge field. It is therefore perfectly possible for the hole dispersion to deviate from the naive free chargon expression, and the fact that the electron dispersion actually agrees with it is a highly non-trivial result which could potentially be relevant for the electron-doped cuprate superconductors. In particular, there is experimental evidence that in the electron-doped cuprates the reconstructed Fermi surface with electron pockets near $(\pi,0)$ and $(0,\pi)$ survives to doping values where the long-range anti-ferromagnetic order has already disappeared~\cite{He2019}. The doped U(1) spin liquid as described by the rotating-frame theory is thus a promising candidate to explain the physics at these doping values, as we have shown that it indeed leads to pockets at the same locations in the Brillouin zone as a doped anti-ferromagnet.

For the bilayer model studied in this work we find that the fermions are always gapped, which means that the U(1) gauge field will be confining for any non-zero monopole fugacity. In a single fermion-layer system --which has a sign problem but is the case relevant for the cuprate superconductors-- the strong inter-layer pairing cannot occur, and there is the possibility of stabilizing a metallic state away from half filling (possibly after adding a repulsive interaction between the fermions). This metallic state could then potentially host a deconfined phase which is stable in the presence of monopoles. But even if the metallic deconfined phase were not stable, the confinement length scale could still be very large such that the physics would be hard to distinguish from that of the doped U(1) spin liquid, especially at finite temperatures. An interesting question for future work is then if and how d-wave superconductivity would emerge from this fractionalized metal.

\begin{acknowledgements}
We thank Zi-Yang Meng and Pietro Bonetti for helpful and stimulating discussions. This research was supported by the European Research Council under the European Union Horizon 2020 Research and Innovation Programme via Grant Agreement No. 101076597-SIESS (X.Z. and N.B.). The computational resources and services used in this work were partially provided by the VSC (Flemish Supercomputer Center), funded by the Research Foundation
- Flanders (FWO) and the Flemish Government. We acknowledge EuroHPC Joint Undertaking for awarding us access to MareNostrum5 hosted by Barcelona Supercomputing Center, Spain.
\end{acknowledgements}

\bibliographystyle{apsrev4-2}
\bibliography{FCP1_QMC}

\onecolumngrid
\appendix
\section{Further details on the QMC simulations}

The Euclidean space-time lattice partition function for the electron-boson model is given by
\begin{equation}
	Z = \int_{\text{C}}[D \bS]\,e^{{-\Delta\tau K\sum_{\Box\perp \tau}f_{\Box}^2}} Z_f^{-1}[\bS]\Tr(\prod_{\tau=1}^{N_\tau}e^{\Delta\tau(\sum_{i,j,l,s} c^{\dagger}_{i,l,s} h_{i,j} c_{j,l,s}+g \sum_{i} (-1)^{x_i+y_i} c^{\dagger}_{i} l^{z}\mathbf{\sigma} c_{i} \cdot \mathbf{S}_{i}(\tau))}),
\end{equation}
where as explained in the main text, the integral over the unit vectors $\bS$ is in a constrained, i.e. hedgehog-free, phase space. The inverse temperature is $\beta=N_\tau \Delta\tau$, and $Z_f^{-1}[\bS]$ is the boson counterterm which cancels the fermion determinant. We have used $l,s$ to label layer and spin, and the free fermion Hamiltonian is
\begin{equation}
h_{i,j}=-t[(\delta_{j,i+\Delta x}+\delta_{j,i+\Delta y})+h.c.]-t'[(\delta_{j,i+\Delta x+\Delta y}+\delta_{j,i+\Delta x-\Delta y})+h.c.]+\mu \delta_{i,j}.
\end{equation}
This bilayer model has a Kramers anti-unitary symmetry $\mathcal{T}=l^x\sigma^y \mathcal{K}$, with $\mathcal{K}$ representing complex conjugation, which ensures that the fermion determinant is positive (and hence the model is sign problem-free). As a result, the bosonic theory including the counter term is hermitian (and local because the fermions are always gapped in the bilayer model). The results of Ref.~\cite{zhang2025constraints} imply that next to anti-ferromagnetism the leading symmetry-breaking channel of the model corresponds to superconductivity with inter-layer spin-singlet electron pairing, i.e. with an order parameter of the form $c^\dagger l^x \sigma^y c^{\dagger}$. In our simulations we fix $\Delta\tau=0.1$ and use a local spin-flip update for the bosonic field. 

\section{Polarization along $\tau$ direction from coupling fermion}

Here we calculate the fermion determinant for the half-filled case with zero hopping. For convenience, we only compute one layer -- the determinant from the other layer is obtained via hermitian conjugation. 

Since the hopping is zero, the determinant factorizes over different lattice sites. The contribution from a single site is given by
\begin{eqnarray}
	&&\Tr(\prod_{\tau=1}^{N_\tau}e^{\Delta \tau g\, \bS(\tau)\cdot c^\dagger(\tau) \boldsymbol{\sigma}c(\tau)}) \\
	&=&2+\sum_{\{n_\tau=\pm1\}}\prod_{\tau=1}^{N_\tau}[\langle n_{\tau}\mathbf{S}(\tau)| n_{\tau+1} \mathbf{S}(\tau+1)\rangle e^{\Delta\tau g n_{\tau}}] \nonumber\\
	&=&2+\sum_{\{n_\tau=\pm1\}} e^{\mathrm{i} F_{n\mathbf{S}}}\prod_{\tau=1}^{N_{\tau}}\frac{|n_{\tau}\mathbf{S}(\tau)+n_{\tau+1}\mathbf{S}(\tau+1)|}{2}e^{\Delta\tau g n_{\tau}}.
\end{eqnarray}
Here, the first term comes from the empty and doubly occupied states, and in the second line we have chosen the quantization axis of the fermion spin to be $\pm\mathbf{S}(\tau)$. We have defined the Wess-Zumino action capturing the Berry phase as $e^{\mathrm{i} F_{n\mathbf{S}}}=e^{\frac{\mathrm{i}}{2} \sum_{\tau=1}^{N_\tau} \Omega_{0,\tau,\tau+1}}$, where $\Omega_{0,\tau,\tau+1}$ is the rotation angle of a tangent vector on spin $S^2$ via the parallel transport starting from a fixed spin $\mathbf{S}_0$, i.e. $\mathbf{S}_0\rightarrow n_{\tau}\mathbf{S}(\tau)\rightarrow n_{\tau+1}\mathbf{S}(\tau+1)\rightarrow \mathbf{S}_0$. $e^{\mathrm{i} F_{n\mathbf{S}}}$ is determined by the rotation angle along imaginary time path of $n_{\tau}\mathbf{S}(\tau)$ and independent with the gauge choice $\mathbf{S}_0$ according to the periodic boundary condition $\mathbf{S}(N_{\tau}+1)=\mathbf{S}(1)$. From the above expression one can see that it is the spin Berry phase which introduces a sign-problem for a single-layer system. The largest term in the sum (obtained by taking all $n_\tau = +1)$ is maximized when the $\bS$ are constant along imaginary-time. The fermion determinant thus introduces a ferromagnetic interaction for the $\mathbf{S}$ along the imaginary time direction. We want to cancel out this effect, and a straightforward way to do this is by adding an additional term to the boson action which exactly cancels this fermion feedback effect. This means that in practice we sample $\mathbf{S}(\tau)$ without considering the contribution from the fermions.

\begin{figure}
	\centering
	\includegraphics[width=0.5\columnwidth]{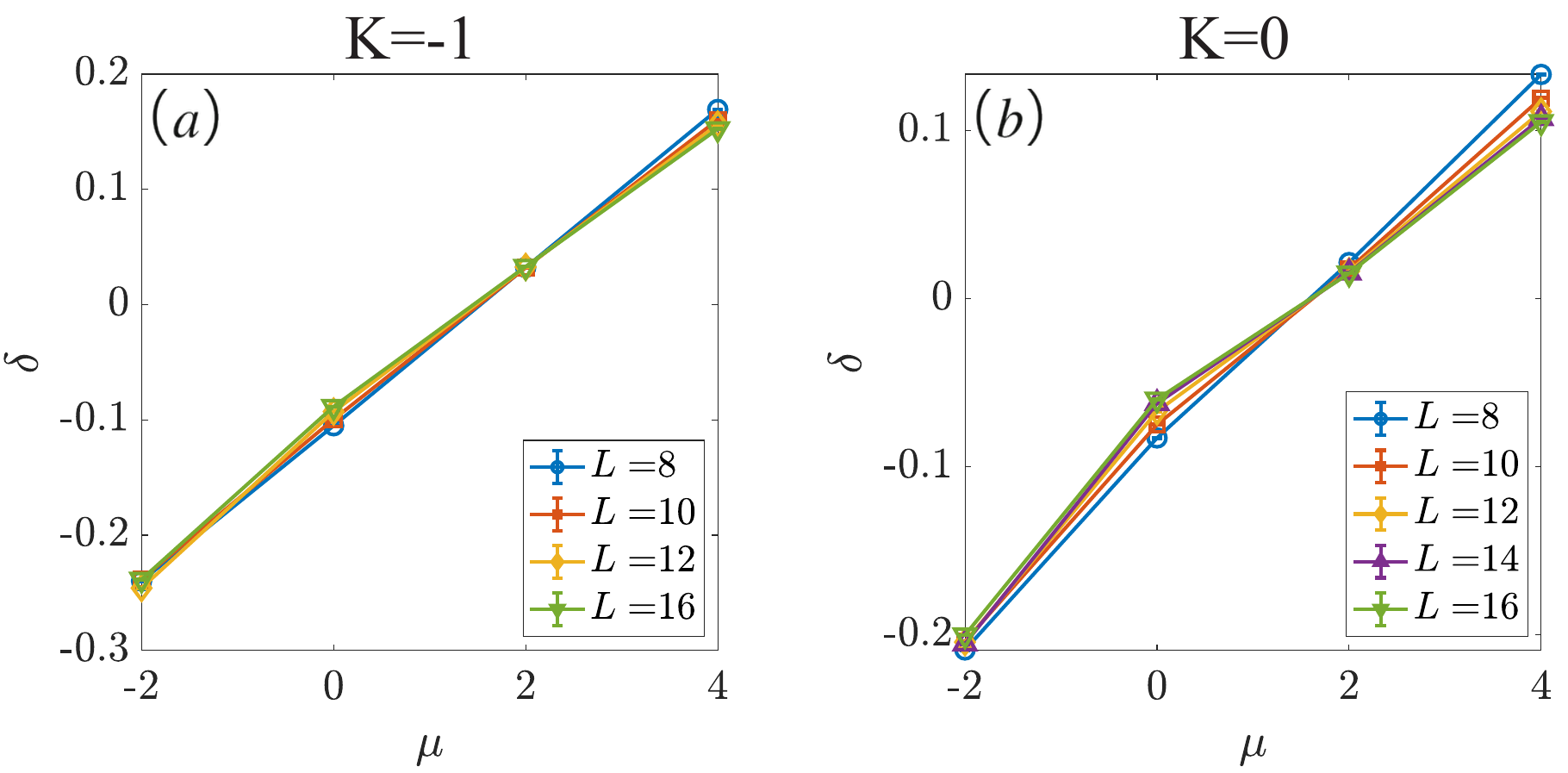}
	\caption{Doping density $\delta$ as a function of chemical potential, obtained with constrained (i.e. hedgehog-free) sampling at both $K=-1$ (a) and $K=0$ (b).}
	\label{fig:Wdelta}
\end{figure} 

\section{Intermediate-coupling results}
In this final appendix we present some numerical results obtained with $t=-3$ and a smaller coupling strength $g=10$. As shown in Fig.~\ref{fig:Wdelta}, for this value of the coupling constant there is no plateau in the doping density $\delta$ as a function of chemical potential (both for $K=-1$ and $K=0$), which shows that the half-filled state is compressible. 

\begin{figure}
	\centering
	\includegraphics[width=1\columnwidth]{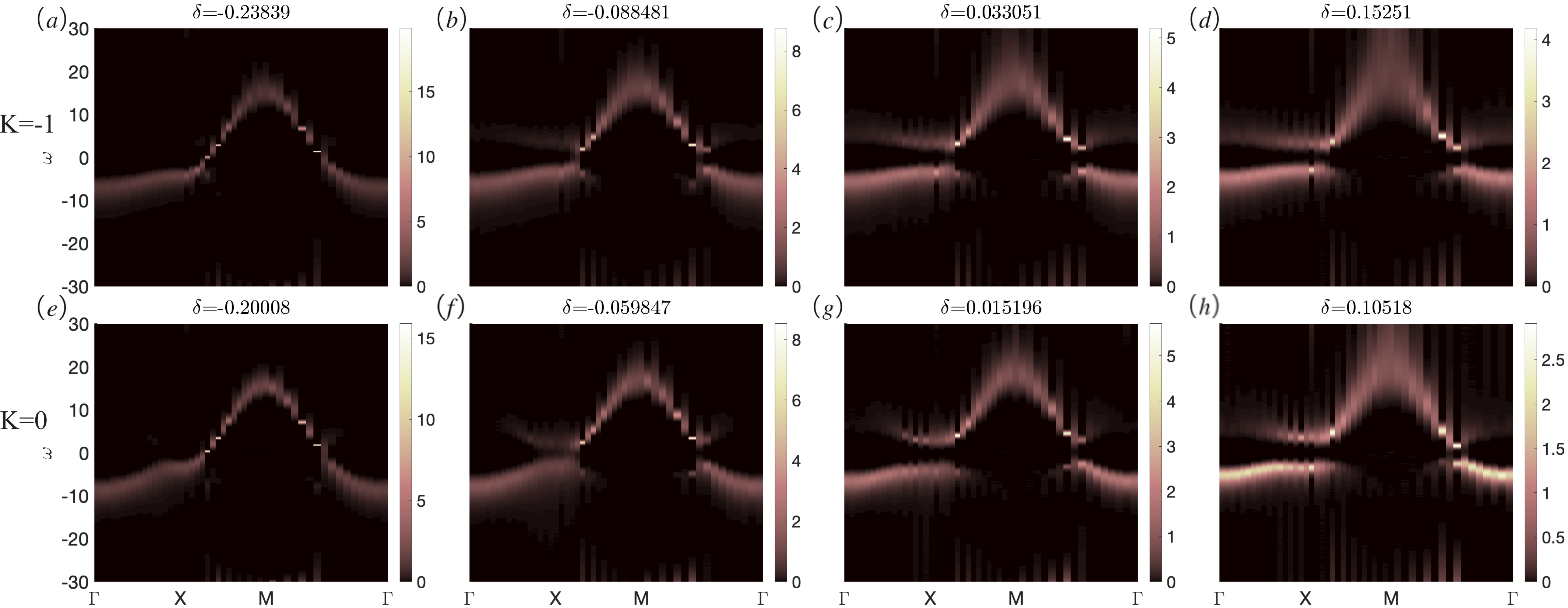}
	\caption{Single-particle spectral weight along a high-symmetry line in the Brillouin zone obtained from stochastic analytical continuation (SAC)~\cite{sandvik1998stochastic,beach2004identifying,shao2023progress} of the imaginary-time Greens' function. Results are obtained with constrained sampling at $K=-1$ (a-d) and $K=0$ (e-h). The single-particle spectrum is gapped in a finite doping range around half filling, indicating a superconducting phase (as the half-filled state is compressible). The system size is $L=L_\tau = 16$, and we have used both periodic and anti-periodic boundary conditions to double the momentum resolution.}
	\label{fig:WSW}
\end{figure}

In Fig.~\ref{fig:WSW} we show the single-particle spectral weight obtained from stochastic analytic continuation of the imaginary-time Green's function~\cite{sandvik1998stochastic,beach2004identifying,shao2023progress} along a high-symmetry line in the Brillouin zone. We see that the fermion spectrum is gapped in a finite doping range around half-filling, indicating superconducting order. 

\begin{figure}
	\centering
	\includegraphics[width=1\columnwidth]{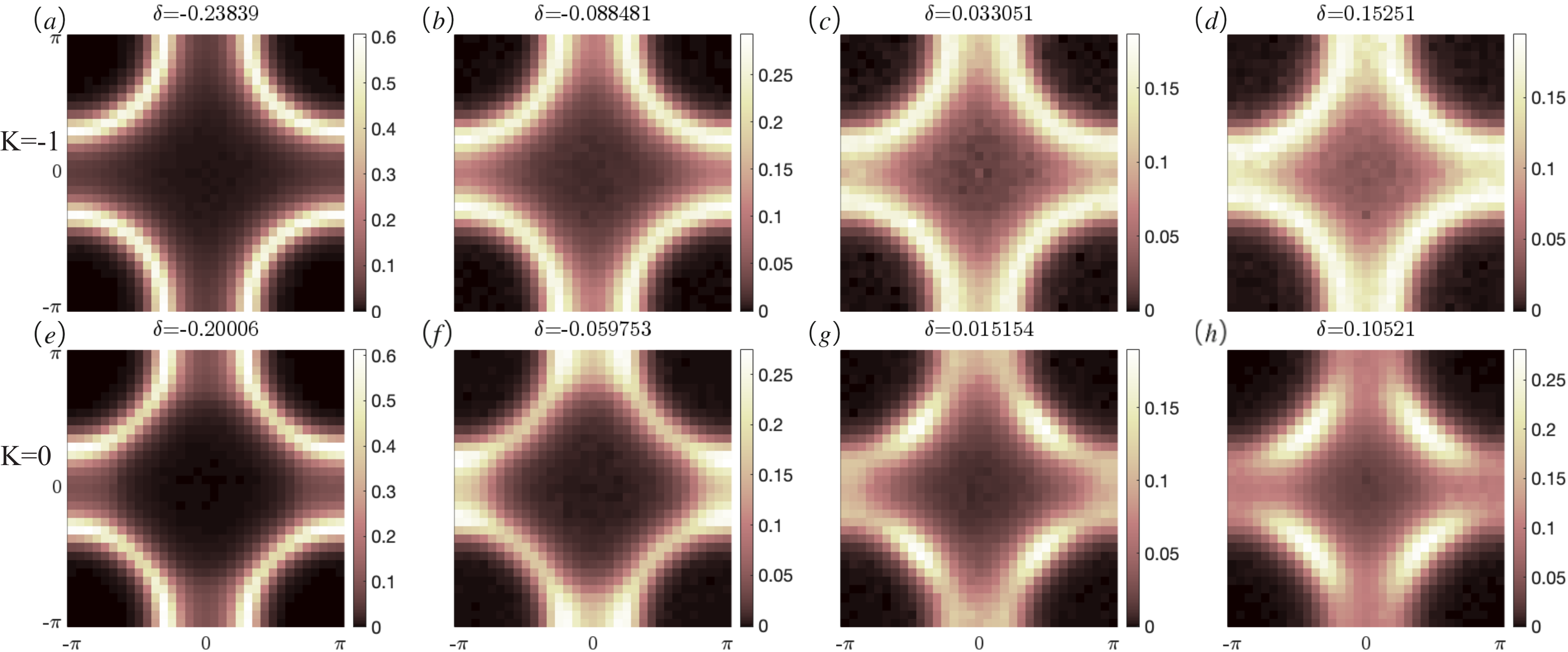}
	\caption{The approximated zero-frequency spectral function from the Green's function as $G(\bk,\tau=\beta/2)\sim A(\bk,0)$ over the Brillouin zone for constrained $K=-1$ (a-d) and $K=0$ (e-h) cases. Though in both cases the system enters the superconduct phase, $K=0$ with a larger AFM fluctuation has shown a fermi-arc shape in (g,h). System size is $L=L_\tau = 16$ and we have used periodic/antiperiodic boundary conditions to enlarge the momentum resolution.}
	\label{fig:WFS}
\end{figure}

In Fig.~\ref{fig:WFS} we show $G(\bk,\tau=\beta/2) = \int_{-\infty}^{+\infty}\mathrm{d}\omega\, A(\bk,\omega)/(2\cosh(\beta \omega))$ as a proxy for the zero-frequency spectral weight $A(\bk,0)$. With $K=-1$ (shown in Figs.~\ref{fig:WFS} (a-d)), for which $\bS$ realizes the symmetric phase of the NCCP$^1$ model, we see that the zero-frequency electronic spectral weight is peaked at the location of the large Fermi surface (i.e. the Fermi surface of the non-interacting theory), as expected when the half-filled state is compressible. For $K=0$ (shown in Figs.~\ref{fig:WFS} (e-h)), where anti-ferromagnetic order develops as the system is cooled to zero temperature, we see that the zero-frequency spectral weight varies significantly along the Fermi surface. On electron-doped side one can see the electron pockets near $(\pi,0)$ and $(0,\pi)$ start to develop, and on the hole-doped side one recognizes the characteristic Fermi arcs near $(\pm \pi/2,\pm \pi/2)$. Interestingly, we find that these features develop at temperatures for which the anti-ferromagnetic correlation length for the fermion spins is very small (order $2$ lattice sites).

\end{document}